\documentclass[11pt,a4paper]{article}
\pdfoutput=1
\usepackage{jheppub}
\usepackage[hyperref,svgnames]{xcolor}
\usepackage[latin1]{inputenc}
\usepackage{amsmath}
\usepackage{amsfonts}
\usepackage{amssymb}
\usepackage{booktabs}
\usepackage{graphicx}
\usepackage{setspace}
\usepackage{mathrsfs}
\usepackage{tikz}
\usepackage{caption}
\usepackage{subcaption}

\definecolor{hgreen}{rgb}{0,.3,0}
\definecolor{hred}{rgb}{.3,0,0}
\definecolor{hblue}{rgb}{0,0,.3}
\definecolor{LightGray}{gray}{0.95}

\numberwithin{equation}{section}

\title{Q-balls with Multiple Charges and Cores\\
}

\author{Olivier Lennon}
\emailAdd{olivier.lennon@physics.ox.ac.uk}

\affiliation{Rudolf Peierls Centre for Theoretical Physics, University of Oxford, Clarendon Laboratory, Parks Road, Oxford OX1 3PU, United Kingdom}

\abstract{Q-balls -- whether in the single-field or multi-field context -- are usually studied in theories containing only one stabilising symmetry. However, this is not the most general scenario. In this paper, we study a class of theories with multiple symmetries. We consider both the traditional thin- and thick-wall limits of these theories, deriving sufficient conditions for existence in the latter case. Moreover, we also introduce a new state that could exist in this class of theory -- a cored Q-ball. We show that this new state can be energetically stable, but leave a detailed phenomenological study to later work.
}

\graphicspath{{./figs/}}

\preprint{OUTP-21-31P}

\begin{document}

\tikzstyle{every picture}+=[remember picture]
\usetikzlibrary{shapes.geometric}
\usetikzlibrary{calc}
\usetikzlibrary{decorations.pathreplacing}
\usetikzlibrary{decorations.markings}
\usetikzlibrary{decorations.text}
\usetikzlibrary{patterns}
\usetikzlibrary{backgrounds}
\usetikzlibrary{positioning}
\tikzstyle arrowstyle=[scale=2]
\tikzstyle directed=[postaction={decorate,decoration={markings,
		mark=at position 0.6 with {\arrow[arrowstyle]{>}}}}]
\tikzstyle rarrow=[postaction={decorate,decoration={markings,
		mark=at position 0.999 with {\arrow[arrowstyle]{>}}}}]

\everymath{\displaystyle}

\maketitle

\section{Introduction}

The states with the lowest mass-to-charge ratio that exist in many theories of complex scalar fields are referred to as Q-balls~\cite{Coleman:1985ki}. These objects are an example of a non-topological soliton~\cite{Lee:1991ax}. These semi-classical objects are kept stable by a combination of energy and Noether charge conservation. Usually, this charge is taken to correspond to be some global $U(1)$ symmetry, but the analysis has been extended to accommodate more complicated symmetry groups~\cite{Safian:1987pr}, or even gauging the symmetry~\cite{Lee:1988ag, Heeck:2021zvk}.

Finding the functional form of these states is a complex process that involves extremising the energy functional of the theory. This is not analytically tractable in the most general of theories. However, analytic limits do exist and can be studied. These are the familiar thin- and thick-wall limits~\cite{Coleman:1985ki, Kusenko:1997ad}, with language borrowed from the mathematically similar study of bounce equations~\cite{Coleman:1977py, Callan:1977pt, Coleman:1977th}. These descriptions are valid in the limits of large and small charge, respectively. Not every theory that possesses an energetically stable thin-wall limit has a stable thick-wall limit~\cite{PaccettiCorreia:2001wtt, Postma:2001ea, Lennon:2021uqu}, implying that there is some gap over which enough charge must accumulate before stability is achieved. Recently, work has been done in Ref.~\cite{Heeck:2020bau} to analytically extend the thin-wall limit to smaller charges.

Q-balls are theoretically interesting objects in their own right. However, they also are interesting phenomenological objects. Due to their stability, they are often considered as candidates for the perceived dark matter in our universe~\cite{Kusenko:1997si, Kusenko:2001vu, Graham:2015apa, Ponton:2019hux}. They could also appear in supersymmetric extensions to the standard model~\cite{Kusenko:1997si, Kusenko:1997zq}, and have been studied in the context of extra dimensions~\cite{Demir:2000gj, Abel:2015tca}. Despite not being yet seen in experiments, it is believed that Q-balls will have striking signals~\cite{Gelmini:2002ez, Kusenko:1997vp, Croon:2019rqu}.

The previous works on single-field theories -- with canonical~\cite{Coleman:1985ki} and non-canonical~\cite{Lennon:2021uqu} kinetic terms -- or multi-field theories -- either coupling a charged field to an arbitrary number of scalars~\cite{Friedberg:1976me,Lennon:2021zzx}, or coupling multiple fields charged under the same symmetry~\cite{Kusenko:1997zq} -- do not represent the most general of theories. One can generalise these ideas further to incorporate multiple symmetries into the theory. Under these circumstances, Q-balls emerge with properties that are different than those from simpler sectors. This work does not represent an exhaustive list of all structures that can exist in this kind of theory, but we hope that it can be the beginning of work towards such a list. One such structure -- a ``barnacled Q-ball'' in the same vein as ``bounces with barnacles'' studied in Refs.~\cite{Balasubramanian:2010kg, Czech:2011aa, Scargill:2017zsz} -- will be returned to in future work.% We merely see this work as an invitation for the community to consider the Q-structures found in more complex theories.

%%%%%%%%%%%%%%%%%%%%

In Section~\ref{sec:MultiFieldMultiSym:Minimisation}, we outline the class of theory of study of this paper. The most straightforward example is a multi-field theory in which each scalar transforms non-trivially under its own symmetry, and is a singlet under all others -- in some sense, this is similar to multiple copies of a single-field theory. The key to this analysis is then the requirement that the potential associated to a single field does not admit Q-ball solutions, and that stability is only achieved due to the couplings between the fields. We then perform the minimisation procedure analogously to the previous Q-ball works. In Section~\ref{sec:MultiFieldMultiSym:ThinWall}, we perform the thin-wall analysis. We determine the differential equation governing the value of the vacuum expectation value (VEV) of the fields inside the resultant Q-ball. Contrary to previous findings, we find that this condition is dependent on the total number of each of the quanta that comprise the Q-ball -- previously, these equations have been charge-independent. In Section~\ref{sec:MultiFieldMultiSym:ThickWall}, we perform the thick-wall analysis, once again making use of the assumption that all fields have the same spatial profile up to a positive semi-definite normalisation constant. Under this assumption, we produce sufficient constraints on the theories that may possess a thick-wall limit. Interestingly, despite the added complexity in the analysis, we find the same result as for single-field theories -- that the next-to-quadratic term must have have an index $p$ such that $2< p < 10/3$. This result certainly adds to the idea that this scenario is similar to multiple copies of a single field theory.

%We close out this part of the thesis with a predominantly heuristic discussion of a new structure within multi-field theories: a cored Q-ball. We envisage a thin-wall Q-ball in one field which possesses a core composed of a thick-wall Q-ball of another field that is stabilised precisely in the homogeneous VEV of the larger Q-ball.\footnote{In principle, these objects are a subclass of ``speckled'' Q-balls, whereby the larger Q-ball is populated by many smaller ones throughout its core -- this is beyond this thesis, but is an interesting avenue of research for the future. This is an interesting addition to the consideration of Q-balls that interact with other fields. This line of thinking has obvious similarities to theories with multiple vacua in the early universe such that the vacuum bubbles develop ``barnacles'' of other vacua -- see Refs.~\cite{Balasubramanian:2010kg, Czech:2011aa, Scargill:2017zsz}.} Each field is stabilised by its own global $U(1)$ symmetry, and so this structure is in principle stable.

In Section~\ref{sec:Cored}, we close out this work with a predominantly heuristic discussion of a new structure within multi-field theories: a cored Q-ball. We envisage a thin-wall Q-ball in one field which possesses a core composed of a thick-wall Q-ball of another field that is stabilised precisely in the homogeneous VEV of the larger Q-ball.\footnote{In principle, these objects are a subclass of ``speckled'' Q-balls, whereby the larger Q-ball is populated by many smaller ones throughout its core -- this is beyond this work, but is an interesting avenue of research for the future. This is an interesting addition to the consideration of Q-balls that interact with other fields. This line of thinking has obvious similarities to theories with multiple vacua in the early universe such that the vacuum bubbles develop ``barnacles'' of other vacua -- see Refs.~\cite{Balasubramanian:2010kg, Czech:2011aa, Scargill:2017zsz}.} Each field is stabilised by its own global $U(1)$ symmetry, and so this structure is in principle stable. We first consider the scenario whereby one field exists in the background VEV of another. We seek Q-ball solutions that are stabilised precisely due to this background VEV. We perform the thick-wall analysis in this scenario, but we note that the calculation proceeds precisely as it would for the single-field case.\footnote{For completeness, we include the thin-wall analysis as an appendix to this paper.} We include these calculations here for completeness and for introducing the notation in the rest of the section. We then consider the case when the background VEV is provided by a large, thin-wall Q-ball. We then merely seek to see whether a cored Q-ball is a stable structure. This leads to two inequalities to be satisfied: the first arises from the requirement that the Q-ball be stable against decay into the individual quanta of both fields; the second arises from the demand that the core be smaller than the Q-ball that houses it. These requirements are model-dependent.

\section{Multi-Charged Q-balls}

We begin this work by analysing the most straightforward example of a multi-field theory with multiple stabilising symmetries -- when each of the fields is charged under its own symmetry.

\subsection{Minimising the Energy in a Sector of Fixed Charge}
\label{sec:MultiFieldMultiSym:Minimisation}

We consider a theory of $N$ complex scalars, $\Phi_i (\vec{x},t)$. The Lagrangian density describing the theory is given by
\begin{equation}
\label{eq:MultiFieldMultiSym:Lagrangian}
\mathcal{L} = \sum^N_i \partial_\mu \Phi_i \partial^\mu \Phi_i^* - U(\Phi_i, \Phi_i^*),
\end{equation}
where $U(\Phi_i, \Phi_i^*)$ is some generic potential that is a function of the fields only, and not their derivatives, and vanishes for vanishing field. The Euler-Lagrange equations for this theory are
\begin{equation}
\partial_{\mu}\partial^{\mu} \Phi_i + \frac{\partial U}{\partial \Phi_i^*} = 0,
\end{equation}
with a similar equation governing the dynamics of $\Phi^*_i$.

We demand that the theory be invariant under an $N$-fold global symmetry,
\begin{equation}
U(1)^{N} = U(1)_1 \times U(1)_2 \times \dots \times U(1)_N,
\end{equation}
that is, $N$ independent $U(1)$ symmetries. We consider here the special case that each field $\Phi_i$ is charged under its own independent $U(1)$ symmetry and is uncharged under all other $U(1)$ symmetries. For each $U(1)$, its corresponding field transforms as
\begin{equation}
\Phi_i \to e^{i q_{i} \alpha} \Phi_i,
\end{equation}
where $\alpha\in\mathbb{R}$ and $q_{i}$ is the charge of the $i$-th species of complex scalar under the $i$-th $U(1)$ symmetry. Associated to each $U(1)$ symmetry is a Noether current density given by
\begin{equation}
\label{eq:MultiFieldMultiSym:MultiNoether}
j^{\mu}_i = i q_{i} \left( \Phi_i \partial^{\mu} \Phi_i^{*} - \Phi_i^{*} \partial^{\mu} \Phi_i \right).
\end{equation}
This symmetry places the constraint on the potential that it be a function of the absolute value of the individual fields.
%\begin{equation}
%U(\Phi_i,\Phi_i^*) = U(\Phi_i^*\Phi_i).
%\end{equation}
We further demand that this potential does not support Q-balls if all but one of the fields vanish: otherwise, this system reduces to that of the single-field case as first discussed in Ref.~\cite{Coleman:1985ki}. We thus assume that the field content is the minimal set of fields, greater than a single field, required to support a Q-ball state.

A Q-ball is the state in a theory which minimises the energy per unit charge. The Hamiltonian for this theory is
\begin{equation}
H = \int \mathrm{d}^3x \left[ \sum^N_i \left(\dot{\Phi}_i \dot{\Phi}_i^* + \vec{\nabla} \Phi_i \cdot \vec{\nabla} \Phi_i^* \right) + U(\Phi_i, \Phi_i^*)\right].
\end{equation}
To determine if this system admits Q-ball solutions, we introduce a set of Lagrange multipliers, $\{\omega_k\}$, similar to Ref.~\cite{Kusenko:1997zq}, that each enforce charge conservation upon minimisation with respect to them:
\begin{equation}
\mathcal{E}_\omega = H + \sum_k^K \omega_k \left( Q_k - \int\mathrm{d}^3 x\, j^0_k \right),
\end{equation}
where $j^0_k$ is the zeroth component of the Noether current density associated to the $k$-th $U(1)$ symmetry given in Eq.~\eqref{eq:MultiFieldMultiSym:MultiNoether}. Thus, the functional we wish to analyse is given by
\begin{equation}
\begin{split}
\mathcal{E}_\omega = & \sum_i^N \omega_i Q_i\\
& + \int\mathrm{d}^3x  \left[ \sum^N_i \left(\dot{\Phi}_i \dot{\Phi}_i^* - i \omega_i q_{i} \left( \Phi_i  \dot{\Phi}_i^{*} - \Phi_i^{*}  \dot{\Phi}_i \right) + \vec{\nabla} \Phi_i \cdot \vec{\nabla} \Phi_i^* \right) + U(\Phi_i, \Phi_i^*) \right].
\end{split}
\end{equation}
We may complete the square on the first two terms under the integral to give
\begin{equation}
\begin{split}
\mathcal{E}_\omega = & \sum_i^N \omega_i Q_i \\
& + \int\mathrm{d}^3x  \left[ \sum^N_i \left(\left|\dot{\Phi}_i - i \omega_i q_{i} \Phi_i \right|^2  + \vec{\nabla} \Phi_i \cdot \vec{\nabla} \Phi_i^* - \omega_i^2 q_{i}^2 \Phi_i^* \Phi_i \right)+ U(\Phi_i, \Phi_i^*) \right].
\end{split}
\end{equation}
The terms containing derivatives with respect to time are the only terms with explicit time dependence. These are positive semi-definite and are minimised if they vanish. Thus, we require, for each species of scalar, that
\begin{equation}
\Phi_i (\vec{x}, t) = e^{i \omega_i q_{i} t} \phi_i (\vec{x}),
\end{equation}
where $\phi_i (\vec{x})$ are functions purely of the spatial coordinate which we take, without loss of generality, to be real-valued. The energy functional is then
\begin{equation}
\label{eq:MultiFieldMultiSym:EnergyFunc}
\mathcal{E}_\omega = \sum_i^N \omega_i Q_i + \int\mathrm{d}^3x  \left[ \sum^N_i \left( \vec{\nabla} \phi_i \cdot \vec{\nabla} \phi_i - \omega_i^2 q_{i}^2 \phi_i^2 \right)+ U(\phi_i) \right].
\end{equation}
Notice that spatial profiles for the $\phi_i$ that vanish everywhere lead to a configuration of zero charge. Thus, a configuration of non-zero charge must have spatial profiles for the charged fields that differs from zero in some finite domain. The spatial profiles satisfy the equations
\begin{equation}
\nabla^2 \phi_i = \frac{1}{2} \frac{\partial}{\partial \phi_i} \left(U(\phi_i) - \omega_i^2 q_i^2 \phi_i^2\right).
\end{equation}
These differential equations take the form of a bounce equation~\cite{Coleman:1977py, Callan:1977pt, Coleman:1977th}, under the potentials given by $U(\phi_i,\psi_i) - \omega^2 q_i^2 \phi_i^2$. The lowest energy configurations are known to be spherically symmetric~\cite{{Coleman:1977th}}, under the boundary conditions of the field vanishing at spatial infinity, and being constant at the coordinate origin, and the first derivative vanishing at spatial infinity and the coordinate origin.

These coupled differential equations cannot, in general, be solved analytically. However, under certain assumptions, analytic progress can be made. Namely, in analogy with the discussion in Ref.~\cite{Kusenko:1997ad}, for stable Q-ball solutions to be found, we require that $\omega_{i,0} \leq \omega_i < m_i/q_i$, where $m_i$ is the mass of the $i$-th complex scalar. The thin-wall limit corresponds to the limit that $\omega_i = \omega_{i,0}$ -- in fact, the thin-wall limit defines $\omega_{i,0}$. The thick-wall limit corresponds to the limit $q_i\omega_i \to m_i^-$.

\subsection{Thin-Wall Q-balls}
\label{sec:MultiFieldMultiSym:ThinWall}

Q-balls in the thin-wall limit are characterised by a spherical, homogeneous core and a thin-shell through which the field interpolates between its core and vacuum values. The Q-ball properties are then well-approximated by those of the core. In this regime, we have that
\begin{equation}
\mathcal{E}_{\omega} \approx \sum_i^N \omega_i Q_i + V\left[ U(\phi_i) - \sum^N_i \omega_i^2 q_{i}^2 \phi_i^2 \right].
\end{equation}
Minimisation of this expression with respect to each of the $\omega_k$ yields
\begin{equation}
Q_k = 2\omega_kVq_k^2\phi_k^2,
\end{equation}
as would be expected from the Noether current density given in Eq.~\eqref{eq:MultiFieldMultiSym:MultiNoether}. Eliminating the Lagrange multipliers from the energy yields
\begin{equation}
E = \frac{1}{4V} \sum_i^N \frac{Q_i^2}{q_i^2 \phi_i^2} + U(\phi_i)V.
\end{equation}
Minimisation with respect to the volume gives us that
\begin{equation}
V^2 = \frac{1}{4 U(\phi_i)} \sum_i^N \frac{Q_i^2}{q_i^2 \phi_i^2},
\end{equation}
which, upon substitution, finally gives us that
\begin{equation}
m_Q = \sqrt{U(\phi_i) \sum_i^N \frac{Q_i^2}{q_i^2 \phi_i^2}}.
\end{equation}
This expression must be minimised with respect to the field content subject to the condition
\begin{equation}
m_Q < \sum_i^N \frac{Q_i}{q_i}m_i,
\end{equation}
where $m_i$ are the masses of the quanta associated to the field $\phi_i$. This ensures that these Q-balls are classically stable against evaporation into the field content. The result of this minimisation is the constraint on the potential:
\begin{equation}
\frac{\partial U(\phi_i)}{\partial \phi_k} \sum_i^N \frac{Q_i^2}{q_i^2 \phi_i^2} = 2 \frac{Q_k^2}{q_k^2 \phi_k^3} U(\phi_i).
\end{equation}
Since this must be true for all $k$, we therefore find that
\begin{equation}
\frac{\partial U(\phi_i)}{\partial \phi_j} \frac{q_j^2 \phi_j^3}{Q_j^2} = \frac{\partial U(\phi_i)}{\partial \phi_k} \frac{q_k^2 \phi_k^3}{Q_k^2}.
\end{equation}

Note, for the single field case -- for a field with unit charge -- the rest mass of a Q-ball is given simply by
\begin{equation}
m_Q = Q \sqrt{\frac{U(\phi)}{\phi^2}},
\end{equation}
where $\phi$ satisfies
\begin{equation}
\frac{\mathrm{d}U(\phi)}{\mathrm{d}\phi} = 2 \frac{U(\phi)}{\phi}.
\end{equation}
In the single-field case, the VEV of the field inside the core is independent of the charge of the Q-ball. As the field has unit charge, this means that it is independent of the total number of quanta which make up the Q-ball.\footnote{This is also true in the single-symmetry multi-field case, both when there is only a single charged scalar~\cite{Lennon:2021zzx} and when there are multiple fields charged under the same symmetry~\cite{Kusenko:1997zq}.} However, in the multi-symmetry case above, we note that the VEVs of the fields inside the Q-ball \textit{do} depend on the total number of quanta (as $N_i = Q_i/q_i$). This is in stark contrast to the known Q-ball cases. As the mass of the core does not simply scale with particle number, we must conclude that it is not composed of ordinary Q-matter.

\subsection{Thick-Wall Q-balls}
\label{sec:MultiFieldMultiSym:ThickWall}

We now consider the thick-wall limit of this theory. This corresponds to the small-field limit of the field theory, as discussed at length in Ref.~\cite{Kusenko:1997ad}. Thus, we will perform a small-field expansion on the function denoting the potential to proceed analytically.

As in the multi-field case with only a single symmetry~\cite{Postma:2001ea, Bishara:2017otb, Lennon:2021zzx}, this theory suffers from the fact that it is not analytically tractable due to the number of fields. As in earlier works, this problem must be circumvented by assuming that all the spatial profiles of the fields are the same up to some semi-positive-definite normalisation constant -- we must then minimise the energy with respect to each of these normalisation constants. We thus write that
\begin{equation}
\phi_i = \alpha_i \phi,
\end{equation}
where $\phi \in \{ \phi_i \}$ is some reference field -- it does not affect the analysis in this case which field is used. The purpose of this ansatz is to provide sufficient conditions for the existence of thick-wall Q-balls in this class of multi-field theories, as well as to provide approximate values for the Q-ball properties. In reality, the spatial profiles of the fields might differ, but this extra freedom in the minimisation process can only further lower the Q-ball energy. To provide necessary conditions for the thick-wall limit, we must pursue a dedicated numerical analysis of the true spatial profiles of all individual fields. Since we have effectively reduced our multi-field system into one of a single-field, we expect to find results similar to those in Ref.~\cite{PaccettiCorreia:2001wtt, Postma:2001ea, Lennon:2021uqu}. We will thus be unsurprised when this does indeed happen below.

Given this approximation scheme, the full bounce potential can be written in the small field limit as
\begin{equation}
U_{\Omega}(\phi, \alpha_i) \approx \left[ \sum_i^N \alpha_i^2 m_i^2 (1 - \Omega_i^2) \right]\phi^2 - g(\alpha_i) \phi^p,
\end{equation}
where $p>2$ and we have defined
\begin{equation}
\Omega_i \equiv \frac{q_i \omega_i}{m_i},
\end{equation}
with $\Omega_i \in (0,1)$, such that the coefficient of the quadratic term is always positive. The function $g(\alpha_i)$ is model-dependent and we leave it general apart from the fact that it must be positive definite for the potential to allow Q-ball solutions. We have omitted the next term in the series, but it is assumed that it is positive definite and stabilises the potential at high field values. The energy in Eq.~\eqref{eq:MultiFieldMultiSym:EnergyFunc} is then
\begin{equation}
\mathcal{E}_\omega = \sum_i^N m_i \Omega_i \frac{Q_i}{q_i} + \int\mathrm{d}^3x  \left[ \sum^N_i \alpha_i^2 \left( \vec{\nabla} \phi \cdot \vec{\nabla} \phi + m_i^2 ( 1 - \Omega_i^2 ) \phi^2 \right) - g(\alpha_i)\phi^p \right].
\end{equation}
We wish to render the integral dimensionless. To do so, we define the dimensionless variables
\begin{equation}
\varphi \equiv \phi \left[ \dfrac{g(\alpha_i)}{\sum_i^N \alpha_i^2 m_i^2 (1 - \Omega_i^2)} \right]^{1/(p-2)} \quad \mathrm{and} \quad \xi_i \equiv x_i \left[ \dfrac{\sum_i^N \alpha_i^2 m_i^2 (1 - \Omega_i^2)}{\sum_i^N \alpha_i^2} \right]^{1/2}.
\end{equation}
The functional of study is then given by
\begin{equation}
\label{eq:MultiFieldMultiSym:EnergyFuncThick}
\mathcal{E}_\omega = \sum_i^N m_i \Omega_i \frac{Q_i}{q_i} + \frac{S_\varphi}{g(\alpha_i)^{2/(p-2)}}\left[\sum_i^N \alpha_i^2\right]^{3/2} \left[ \sum_i^N \alpha_i^2 m_i^2 ( 1 - \Omega_i^2 ) \right]^{(6-p)/(2p-4)},
\end{equation}
where $S_\varphi$ is a dimensionless integral, defined as
\begin{equation}
S_{\varphi} = \int\mathrm{d}^3\xi \left[ \vec{\nabla}_\xi \varphi \cdot \vec{\nabla}_\xi \varphi + \varphi^2 - \varphi^p \right].
\end{equation}
For different values of $p$, this has been numerically minimised in Ref.~\cite{Linde:1981zj}, with the general trend being that $S_{\varphi}$ increases for increasing $p$. This expression must be minimised with respect to the $\{\Omega_i\}$ and $\{\alpha_i \}$. Without specifying the model, it is difficult to make progress. However, as in previous work~\cite{PaccettiCorreia:2001wtt, Postma:2001ea, Lennon:2021uqu, Lennon:2021zzx}, we can place limits on the theories that can house Q-ball states.

We minimise the above function in Eq.~\eqref{eq:MultiFieldMultiSym:EnergyFuncThick} with respect to some $\Omega_k$. This yields the condition
\begin{equation}
\frac{Q_k}{q_k \alpha_k^2 m_k \Omega_k} = \frac{S_\varphi}{g(\alpha_i)^{2/(p-2)}} \left[\frac{6-p}{p-2}\right] \left[\sum_i^N \alpha_i^2\right]^{3/2} \left[ \sum_i^N \alpha_i^2 m_i^2 ( 1 - \Omega_i^2 ) \right]^{(10-3p)/(2p-4)}.
\end{equation}
Notice, the left hand-side is $k$-dependent, but the right-hand side is not. Since this condition holds for all $k$, we can readily write
\begin{equation}
\label{eq:MultiFieldMultiSym:ThickMinCondition}
\frac{Q_k}{q_k \alpha_k^2 m_k \Omega_k} = \frac{Q_i}{q_i \alpha_i^2 m_i \Omega_i}
\end{equation}
for all $k,i\in\{1,\cdots,N\}$. Given that $\alpha = 1$ for, say, the field $\phi_1$, we have that
\begin{equation}
\alpha_k^2 = \frac{q_1}{Q_1} \frac{Q_k}{q_k} \frac{m_1}{m_k} \frac{\Omega_1}{\Omega_k},
\end{equation}
and so $\alpha$ is now defined for all fields in the theory.

Using our above condition for minimisation, we can rewrite Eq.~\eqref{eq:MultiFieldMultiSym:EnergyFuncThick} as
\begin{equation}
\mathcal{E}_\omega = \sum_i^N m_i \Omega_i \frac{Q_i}{q_i} + \frac{Q_k}{q_k \alpha_k^2 m_k \Omega_k} \left[\frac{p-2}{6-p}\right] \left[\sum_i^N \alpha_i^2\right]^{3/2} \left[ \sum_i^N \alpha_i^2 m_i^2 ( 1 - \Omega_i^2 ) \right].
\end{equation}
At its minimum, this expression must satisfy
\begin{equation}
\mathcal{E}_\omega < \sum_i^N \frac{Q_i}{q_i}m_i
\end{equation}
in order for the resulting Q-balls to be classically stable against decay to the constituent quanta of the complex scalar fields. From Eq.~\eqref{eq:MultiFieldMultiSym:ThickMinCondition}, we see that we can write
\begin{equation}
\sum_i^N m_i \Omega_i \frac{Q_i}{q_i}  = \frac{Q_k}{q_k \alpha_k^2 m_k \Omega_k} \sum_i^N \alpha_i^2 m_i^2 \Omega_i^2,
\end{equation}
and
\begin{equation}
\sum_i^N \frac{Q_i}{q_i}m_i = \frac{Q_k}{q_k \alpha_k^2 m_k \Omega_k} \sum_i^N \alpha_i^2 m_i^2 \Omega_i.
\end{equation}
Thus, our condition on the minimum becomes
\begin{equation}
\sum_i^N \alpha_i^2 m_i^2 \left[ \Omega_i^2 + \left(\frac{p-2}{6-p}\right)(1 - \Omega_i^2) - \Omega_i \right] < 0.
\end{equation}
For this condition to hold, we necessarily require that that term in brackets be negative for some $i$ in the range $\Omega \in (0,1)$. However, this is nothing more than the same requirement on the canonical single-field case -- see Ref.~\cite{Lennon:2021uqu}. We thus infer that, for stable thick-wall Q-balls to form in this theory, we must have that $2 < p < 10/3$.

\section{Cored Q-balls}
\label{sec:Cored}

We now turn our attention to another type of Q-ball that can plausibly exist in multi-field theories with multiple global symmetries. We refer to these as ``cored Q-balls''. In this section, we merely show that these objects can be classically stable against decay into constituent quanta.

\subsection{Q-balls in the Background of Another Field}
\label{sec:CoredQballs:Background}

A key component of this section is the idea of a Q-ball stabilised in the background of another field. We therefore formulate this idea for a background VEV extending over all space.

\subsubsection{The Minimisation Procedure}

Consider the $N=2$ case of the Lagrangian given in Eq.~\eqref{eq:MultiFieldMultiSym:Lagrangian}. We write the potential of this theory as
\begin{equation}
U(\Phi_1,\Phi_2) = f(\Phi_1) + g(\Phi_2) + h(\Phi_1, \Phi_2),
\end{equation}
where it is understood that $h$ contains all terms that couple the two fields together. We assume that the field $\Phi_1$ acquires some constant VEV -- in this subsection, we leave the method of acquisition of this VEV open, but in the next subsection, we will assume that it comes after the formation of a thin-wall Q-ball. We thus seek Q-ball solutions for the field $\Phi_2$ in this constant background. Furthermore, this constant background will merely change the coefficients of the field $\Phi_2$ by constant amounts, and so this analysis will proceed exactly as for a single-field system. For definiteness, we perform the analysis below.

The energy of a configuration of $\Phi_2$ in the background of $\bar{\phi}_1$ is
\begin{equation}
H = \int\mathrm{d}^3x \left[ \dot{\Phi}_2 \dot{\Phi}_2^* + \vec{\nabla}\Phi_2 \cdot \vec{\nabla} \Phi_2^* + g(\Phi_2) + h(\bar{\phi}_1, \Phi_2) \right].
\end{equation}
A Q-ball is a state that minimises the energy for a given charge and so we introduce a Lagrange multiplier, $\omega_2$, that ensures charge conservation upon minimisation with respect to it. The functional we wish to analyse is then
\begin{equation}
\mathcal{E}_\omega = H + \omega_2 \left(Q_2 - \int\mathrm{d}^3x \, j_2^0 \right),
\end{equation} 
where $j_2^0$ is the zeroth component of the Noether current density associated to the $U(1)_2$ symmetry,
\begin{equation}
j_2^0 = i q_2 \left( \dot{\Phi}_2^* \Phi_2 - \Phi_2^* \dot{\Phi}_2 \right).
\end{equation}
We may thus rewrite our functional of study as
\begin{equation}
\begin{split}
\mathcal{E}_\omega = & \int\mathrm{d}^3x \left[ \left| \dot{\Phi}_2 - i \omega_2 q_2 \Phi_2 \right|^2 + \vec{\nabla}\Phi_2 \cdot \vec{\nabla} \Phi_2^* + g(\Phi_2) + h(\bar{\phi}_1, \Phi_2) - \omega_2^2 q_2^2 \Phi_2^* \Phi_2 \right]\\
& + \omega_2 Q_2.
\end{split}
\end{equation}
Note, the term containing explicit time-dependence is positive semi-definite. This is therefore minimised when it vanishes, i.e., if
\begin{equation}
\Phi_2(\vec{x},t) = e^{i \omega_2 q_2 t} \phi_2(\vec{x}),
\end{equation}
where we take the spatial profile to be real-valued, without loss of generality. Reinsertion of this into the functional of study leads to
\begin{equation}
\mathcal{E}_\omega = \omega_2 Q_2 +\int\mathrm{d}^3x \left[ \vec{\nabla}\phi_2 \cdot \vec{\nabla} \phi_2 + g(\phi_2) + h(\bar{\phi}_1, \phi_2) - \omega_2^2 q_2^2 \phi_2^2 \right].
\end{equation}
Minimisation with respect to the spatial profile yields a differential equation governing a ``bounce'' solution associated to the formation of vacuum bubbles during phase transitions~\cite{Coleman:1977py, Callan:1977pt, Coleman:1977th}. These equations are well-studied and, for a set of boundary conditions appropriate for Q-balls, are known to yield spherically symmetric solutions for the lowest energy configurations.

\subsubsection{The Thick-Wall Limit}

We now pursue the thick-wall limit of this analysis, which is also known as the small field limit.\footnote{We note that ``small-field'' here only relates to $\phi_2$, as it is from this that this Q-ball is being formed.} In principle, we could also form thin-wall Q-balls in the background of another field -- this will not be relevant to the discussion of cored Q-balls, but we include the analysis in Appendix~\ref{app:ThinWall} for completeness.

To proceed in the thick-wall limit, we expand $\phi_2$ to its lowest order terms. As stated above, this analysis is essentially identical to a single field case and so is bound by the requirement that the next-to-quadratic term must be cubic in the bounce potential -- see Refs.~\cite{PaccettiCorreia:2001wtt, Postma:2001ea, Lennon:2021uqu}. We thus assume that this term exists. We therefore have, in the small-field limit, that
\begin{equation}
g(\phi_2) + h(\bar{\phi}_1, \phi_2) \approx \left( \mu^2 + h_2 (\bar{\phi}_1) \right) \phi_2^2 - \left( h_3(\bar{\phi}_1) - A \right) \phi_2^3,
\end{equation}
where $h_2(\bar{\phi}_1)$ and $h_3(\bar{\phi}_1)$ are the components, functionally dependent on $\bar{\phi}_2$, of the terms quadratic and cubic in $\phi_2$ contained within $h(\bar{\phi}_1, \phi_2)$, respectively, and similarly for $\mu^2$ and $A$ in $g(\phi_2)$. We assume that some positive definite term of a higher order exists to stabilise the potential at large field. In this scenario, a Q-ball may only form if the coefficients of the quadratic and cubic terms are positive and negative, respectively:
\begin{equation}
\mu^2 + h_2 (\bar{\phi}_1) >  0 \quad \mathrm{and} \quad h_3(\bar{\phi}_1) - A >  0.
\end{equation}
However, recall that we demanded that the pure $\Phi_2$ potential could not support Q-ball solutions. This sets $A \geq 0$, hence the sign assignment on the cubic term.

For notational purposes, we will write the effective quadratic and cubic couplings in the background as
\begin{equation}
\mu^2_{\mathrm{eff}} (\bar{\phi}_1) \equiv  \mu^2 + h_2 (\bar{\phi}_1) \quad \mathrm{and} \quad A_{\mathrm{eff}} (\bar{\phi}_1) \equiv h_3(\bar{\phi}_1) - A,
\end{equation}
noting that both of these new parameters are positive definite. The bounce potential can now be written in the familiar form,
\begin{equation}
U_{\omega} (\bar{\phi}_1, \phi_2) \approx (\mu_{\mathrm{eff}}^2 - \omega_2^2 q_2^2)\phi_2^2 - A_{\mathrm{eff}}\phi_2^3.
\end{equation}
For a barrier to form in the bounce potential, and thus for a bounce solution to exist -- see Ref.~\cite{Kusenko:1997ad} -- we see that
\begin{equation}
\omega_2^2 < \frac{\mu_{\mathrm{eff}}^2}{q_2^2},
\end{equation}
which is the familiar requirement of Q-ball solutions. The thick-wall limit is thus equivalent to the limit $\omega_2 q_2 \to \mu_{\mathrm{eff}}^{-}$.

We redefine the spatial coordinate and field by choosing
\begin{equation}
\begin{split}
\xi_i = & \left(\mu_{\mathrm{eff}}^2  - \omega_2^2 q_2^2 \right)^{1/2} x_i \\
\psi = & \left(\mu_{\mathrm{eff}}^2  - \omega_2^2 q_2^2 \right)^{-1} A_{\mathrm{eff}} \phi_2.
\end{split}
\end{equation}
Our functional of study thus reduces to
\begin{equation}
\mathcal{E}_\omega = \mu_{\mathrm{eff}}^3\dfrac{\left( 1  - \Omega^2 \right)^{3/2}}{A_{\mathrm{eff}}^2} S_\psi + \omega_2 Q_2,
\end{equation}
where we have defined $\Omega \equiv \omega_2 q_2 /\mu_{\mathrm{eff}}$ and
\begin{equation}
S_\psi = \int\mathrm{d}^3\xi \left[ \vec{\nabla}_\xi \psi \cdot \vec{\nabla}_\xi \psi +\psi^2 - \psi^3 \right]
\end{equation}
is a dimensionless integral that may be minimised numerically. This has been numerically found to be approximately 38.8~\cite{Linde:1981zj}.

A stable Q-ball in the background of $\bar{\phi}_1$ is found when minimising this expression with respect to $\Omega$ under the constraint that $ \Omega^2 < 1$. If this constraint is not satisfied, it is more energetically favourable for the quanta of $\Phi_2$ to remain seperate within the background of $\bar{\phi}_1$. This requirement leads us to the condition
\begin{equation}
\epsilon = \Omega (1 - \Omega^2)^{1/2},
\end{equation}
where we define
\begin{equation}
\label{eq:CoredQballs:ThickEpsilon}
\epsilon \equiv \frac{1}{3 S_{\varphi}}  \frac{Q_2}{q_2} \frac{A^2_{\mathrm{eff}}}{\mu^2_{\mathrm{eff}}}.
\end{equation}
The solution to this condition is that
\begin{equation}
\Omega^2 = \frac{1 + \sqrt{1 - 4 \epsilon^2}}{2},
\end{equation}
where $0 < \epsilon < 1/2$. The mass of the resulting Q-ball is then given by
\begin{equation}
m_Q = \mu_{\mathrm{eff}} \frac{Q_2}{q_2} \left( 1 - \frac{1}{6} \epsilon^2 + \mathcal{O}(\epsilon^4) \right).
\end{equation}
We note that
\begin{equation}
m_Q < \mu_{\mathrm{eff}}\frac{Q_2}{q_2},
\end{equation}
which is precisely the requirement for classical stability of this Q-ball within the background of $\bar{\phi}_1$. The radius of the resulting Q-ball is given by $\xi \sim 1$, and so
\begin{equation}
\label{eq:CoredQballs:ThickRadius}
R^{-1} \sim \epsilon \mu_{\mathrm{eff}} \left(1 + \mathcal{O} \left( \epsilon^2 \right) \right).
\end{equation}

This solution is also subject to the constraint that $\epsilon < 1/2$, which translates to the requirement on the charge that
\begin{equation}
\label{eq:CoredQballs:UpperBoundConstrain-thick}
\frac{Q_2}{q_2} < 58.2 \frac{\mu_{\mathrm{eff}}}{A^2_{\mathrm{eff}}}.
\end{equation}
The solution is also subject to the constraint that the higher order term that stabilises the potential is indeed small enough inside the Q-ball to ignore in our analysis -- see Ref.~\cite{Kusenko:1997ad} for details.

\subsection{Cored Q-balls}
\label{sec:CoredQballs:CoredQballs}

We assume that the potential $f(\Phi_1)$ admits Q-balls composed of the field $\Phi_1$. In the thin-wall limit, these Q-balls are spherically-symmetric, extended objects comprised of a large core of homogeneous ``Q-matter'', and a thin shell which interpolates between the core and the vacuum of the theory. The VEV in the core is therefore the origin of the VEV of $\Phi_1$ in the previous subsection. The physical properties of the Q-ball -- it's mass and volume -- are well-approximated by the properties of the core in this limit. Inside the Q-ball, where the field is homogeneous, $\Phi_1$ takes on the well-known functional form
\begin{equation}
\Phi_1 (t) = e^{i q_1 \omega t} \bar{\phi}_1,
\end{equation}
where $\bar{\phi}_1 \in \mathbb{R}$. The rest mass and volume of the Q-ball are then given by
\begin{equation}
\label{eq:CoredQballs:ThinPhysical}
m_Q = Q_1 \sqrt{\frac{f(\bar{\phi}_1)}{q_1^2 \phi_1^2}} \quad \mathrm{and} \quad V = \frac{Q_1}{2 \sqrt{q_1^2 \bar{\phi}_1^2 f(\bar{\phi}_1)}},
\end{equation}
where $\bar{\phi}_1$ must satisfy
\begin{equation}
\frac{\partial f(\bar{\phi}_1)}{\partial \bar{\phi}_1} = \frac{2 f(\bar{\phi}_1)}{\bar{\phi}_1},
\end{equation}
such that $E/Q_1$ must be less than the mass per unit charge of individual quanta of the $\Phi_1$ field. It is in this way that these Q-balls are said to be classically stable.

We assume that, if anything, $h(\Phi_1, \Phi_2)$ provides, at most, a small perturbation to this Q-ball solution. For definiteness, we assume that $g(\Phi_2)$ does not admit Q-ball solutions in isolation from $\Phi_1$, and we further assume, for simplicity, that $U(\Phi_1, \Phi_2)$ does not admit Q-balls charged under both symmetries when all terms are considered.

Now, consider a configuration where a thin-wall Q-ball composed of $\Phi_1$ is surrounded by quanta of the field $\Phi_2$. The potential $h(\Phi_1, \Phi_2)$ defines how these entities interact. Let us consider an attractive potential such that the quanta of $\Phi_2$ become bound within the Q-ball. We will only consider a small amount entering the Q-ball such that the value of $\bar{\phi}_1$ is still consistent with that defined above. Though the core of the Q-ball is, in principle, homogeneous, the presence of the new quanta slightly breaks this homogeneity, and we expect these to settle in the core of the Q-ball as this is the symmetric point of the Q-ball.\footnote{We could also consider the case that many thick-wall Q-balls form within the homogeneous background of the thin-wall Q-ball. However, we do not consider this ``speckled'' Q-ball here.} %Given the interactions of the quanta of $\Phi_2$ mediated by the Q-matter, we can expect that a thick-wall Q-ball can form in the core, as per the analysis in the previous section, provided that enough quanta congregate within this region.

If enough quanta are collected, it is reasonable to ask whether a Q-ball composed of $\Phi_2$, in the background of $\bar{\phi}_1$, can form. Moreover, if this Q-ball does not get too large, such that the terms in $h(\Phi_1, \Phi_2)$ greatly perturb the Q-ball background, it should not greatly affect the value of $\bar{\phi}_1$. We will assume that the core is a thick-wall Q-ball, as this is more consistent with our assumption that the new Q-ball does not overly affect the value of $\bar{\phi}_1$, since the VEV is smaller. We can take the results directly from the previous section.

We require that the whole system be energetically stable against classical decay to quanta of both fields. The total mass of the cored Q-ball is
\begin{equation}
E_{\mathrm{tot}} = \frac{Q_1}{q_1} \sqrt{\frac{f(\bar{\phi}_1)}{\bar{\phi}_1^2}} + \frac{Q_2}{q_2} \mu_{\mathrm{eff}} \left(1 - \frac{1}{6} \epsilon^2 \right).
\end{equation}
For this to be classically stable against decay to the quanta of both of the fields, we require that
\begin{equation}
E_{\mathrm{tot}} < \frac{Q_1}{q_1} m + \frac{Q_2}{q_2} \mu,
\end{equation}
where $m$ is the mass of quanta of the field $\Phi_1$ and, as before, $\mu$ is the mass of the quanta of the field $\Phi_2$ in the vacuum of the theory, $\Phi_1 = 0$. Note that, the first term of the total energy is less than the first term of this constraint. If $\mu_{\mathrm{eff}} < \mu$, then this constraint is satisfied. If this is not true, then a sufficient condition is for\footnote{The necessary condition is that the the binding energy of the thin-wall Q-ball is greater than any energy cost that the thick-wall Q-ball expanded in the background of $\bar{\phi}_1$ brings over the vacuum of the theory,
\begin{equation}
\frac{Q_1}{q_1} \left(m - \sqrt{\frac{f(\bar{\phi}_1)}{\bar{\phi}_1^2}}  \right) > \frac{Q_2}{q_2} \left[ \mu_{\mathrm{eff}} \left(1 - \frac{1}{6} \epsilon^2 \right) - \mu \right].
\end{equation}
This is a model-dependent condition, so we say nothing more of this here.
}
\begin{equation}
\mu_{\mathrm{eff}}  \left(1 - \frac{1}{6} \epsilon^2 \right) < \mu.
\end{equation}
This constraint, in terms of the charge of the thick-wall Q-ball, evaluates to
\begin{equation}
\frac{Q_2}{q_2} > 201.6 \sqrt{h_2 (\bar{\phi}_1)} \frac{\mu_{\mathrm{eff}}}{A_{\mathrm{eff}}^2}.
\end{equation}
This implies that, for this structure to be classically stable, there is a minimum charge that must accumulate in the thick-wall core. For this to be compatible with Eq.~\eqref{eq:CoredQballs:UpperBoundConstrain-thick}, we require that
\begin{equation}
11h_2 < \mu.
\end{equation}
We thus see that we require that the coupling between the two fields not be too large. If this is satisfied, then it is sufficient to prove that this structure is classically stable over some range of charge. The total system is therefore adequately described as a thin-wall Q-ball, stabilised by the conservation of some charge, with a core composed of a thick-wall Q-ball, stabilised by the conservation of some other charge, over the background field of the thin-wall Q-ball. This is because the field $\phi_2$ will remain ``small'' and thus will not overly affect the background field, $\bar{\phi}_1$.

By definition, a Q-ball is the state that minimises the energy per unit charge of a sector of a theory. Thus, the combination of energy and Noether charge conservation implies that this configuration is more stable than emitting the individual quanta of $\Phi_2$ outside the Q-ball formed from $\Phi_1$. One could also question whether a Q-ball composed of $\Phi_1$ and $\Phi_2$ could be emitted as a decay channel. However, the ability for the potential defined through $g(\Phi_2)$ and $h(\Phi_1, \Phi_2)$ in the background of $\bar{\phi}_1$ to house stable Q-ball solutions does not imply that the full potential $U(\Phi_1,\Phi_2)$ will allow for stable ``doubly-charged'' Q-balls to exist in isolation. In the absence of this decay mode, the cored Q-ball is indeed stable.

We also require that the Q-ball in the core be smaller than the thin-wall Q-ball it resides in, as otherwise we cannot assume the results of the previous section. From the sizes of the respective Q-balls given in Eqs.~\eqref{eq:CoredQballs:ThickRadius}~and~\eqref{eq:CoredQballs:ThinPhysical}, together with the definition for $\epsilon$ given in Eq.~\eqref{eq:CoredQballs:ThickEpsilon}, we find that
\begin{equation}
36\pi (S_{\varphi})^3 \frac{q_2^3}{Q_2^3} \frac{\mu_{\mathrm{eff}}^3}{A_{\mathrm{eff}}^6} \ll \frac{Q_1}{2 \sqrt{q_1^2 \bar{\phi}_1^2 f(\bar{\phi}_1)}},
\end{equation}
which is a model-dependent condition which relates the charge accumulation of both Q-balls.

\section{Summary}

In this work, we studied a subclass of multi-field and multi-symmetry theories for Q-ball solutions. Specifically, we considered a theory of $N$ complex scalar fields, each charged under their own $U(1)$ symmetry. Thus, the stabilising symmetry is an $N$-fold $U(1)$ symmetry. In this scheme, we studied both the thin- and thick-wall limits, using the standard multi-field approximation of similar spatial profiles in order to progress analytically.

In the thin-wall case, we determined the physical properties of the Q-ball, together with the differential equation that governs the VEV of the fields inside the Q-ball. In sharp contrast to all previous studies, we have found that the VEVs depend upon the charge of the Q-ball itself. The differential equation must then be solved for each charge configuration, independently. It would be interesting to solve this in future work in, say, the ``doubly-charged'' case.

In the thick-wall case, we demanded that the minimum of energy be classically stable against decay into constituent quanta of the fields. We thus gave sufficient constraints on the theories that possess a thick-wall limit. The general set up of the theory was such that we had, in some sense, $N$ copies of the single-field case, and so it comes to no surprise that the restriction of theories is the same as in Refs.~\cite{PaccettiCorreia:2001wtt, Postma:2001ea, Lennon:2021uqu}, namely, that $2<p<10/3$, where $p$ denotes the index of the next-to-quadratic term in the bounce potential.

In the latter half of this work, we heuristically studied a new type of object that we called a ``cored Q-ball''. This is a thin-wall Q-ball whose homogeneous interior acts as a stabilising background VEV for a thick-wall Q-ball composed of another field. In this work, we merely concerned ourselves with the question of stability of these objects. Realistic theories that lead to the formation of these objects is a direction for future work. Specifically, a phenomenological analysis is required in which all timescales are taken into account, together with a mechanism for the release of the additional energy. Moreover, one could also extend the ideas here to the case where a stable, ``doubly-charged'' Q-ball can exist. This would open up the decay channel whereby a mixed Q-ball is emitted, leaving behind a pure thin-wall Q-ball.

Intriguingly, these objects could be important in DM experiments. Consider the case that the Q-ball at the core is charged under the SM, whereas the larger thin-wall Q-ball is not directly coupled to the SM. If this object were to pass through a direct detection experiment, we would only see the core, and not the Q-ball at large. This would cause a misidentification of the DM until experiments are sensitive enough to determine the other component, which could indirectly couple to the SM via the cored field.

\section*{Acknowledgments}

OL would like to thank John March-Russell for useful comments on this work, which was completed while OL was supported by the Colleges of St John and St Catherine, Oxford. Conversations with Fady Bishara on Q-ball physics over the years have also been particularly fruitful.

\appendix

\section{Thin-Wall Q-balls in the Background of Another Field}
\label{app:ThinWall}

For completeness, we now include the thin-wall analysis. A thin-wall Q-ball's physical properties are well-described by those of its homogeneous core. We thus write
\begin{equation}
\mathcal{E}_{\omega} \approx \bar{\omega}_2 Q_2 + V \left[ g(\bar{\phi}_2) + h(\bar{\phi}_1, \bar{\phi}_2) - \bar{\omega}_2^2 q_2^2 \bar{\phi}_2^2 \right],
\end{equation}
where $V$ is the volume of the core, and we use bars to represent values of variables within the core. Minimisation of this with respect to the Lagrange multiplier yields, as expected, the expression for the charge of this configuration,
\begin{equation}
Q_2 = 2\bar{\omega}_2 q_2^2 \bar{\phi}_2^2 V.
\end{equation}
Eliminating $\bar{\omega}_2$ and minimising with respect to the volume leads to
\begin{equation}
V^2 = \frac{Q_2^2}{4 q_2^2 \bar{\phi}_2^2} \frac{1}{\left[ g(\bar{\phi}_2) + h(\bar{\phi}_1, \bar{\phi}_2) \right]}.
\end{equation}
Finally, upon elimination of the volume, we obtain the rest mass of the configuration
\begin{equation}
m_Q = Q_2 \sqrt{\frac{g(\bar{\phi}_2) + h(\bar{\phi}_1, \bar{\phi}_2)}{q_2^2 \bar{\phi}_2^2}},
\end{equation}
which is completely expected from the single-field theory, since $U(\bar{\phi}_2) = g(\bar{\phi}_2) + h(\bar{\phi}_1, \bar{\phi}_2)$. For the resultant Q-ball to be stable in the background of the VEV of $\Phi_1$, we must have that
\begin{equation}
m_Q < \mu_{\mathrm{eff}} \frac{Q_2}{q_2},
\end{equation}
where $\mu_{\mathrm{eff}}(\bar{\phi}_1)$ is defined as in the previous section. The VEV of $\bar{\phi}_2$ must obey the differential equation given by
\begin{equation}
\frac{\partial g}{\partial \bar{\phi_2}} + \frac{\partial h}{\partial \bar{\phi_2}} = 2\left(\frac{g}{\bar{\phi_2}} + \frac{h}{\bar{\phi_2}} \right).
\end{equation}
This concludes the thin-wall analysis of this theory.

\bibliographystyle{JHEP}
\bibliography{paper}
\end{document}